\documentclass[superscriptaddress,12pt,onecolumn,nofootinbib]{revtex4-2} 

\usepackage{mathtools}

\usepackage{booktabs,makecell,array}
\newcolumntype{P}[1]{>{\raggedright\arraybackslash}p{#1}}

\usepackage{amsmath}  
\usepackage{amsfonts} 
\usepackage{graphicx} 
\usepackage{esvect}
\usepackage{mathtools}

\usepackage{xcolor}
\usepackage{appendix}
\usepackage{siunitx} 
\usepackage{mathrsfs}

\usepackage{placeins}

\usepackage[caption=false]{subfig} 

\usepackage{pbox}

\usepackage{mathtools}
\makeatletter
\newcommand{\vast}{\bBigg@{3}}
\newcommand{\Vast}{\bBigg@{3.5}}
\makeatother

\begin{document}


\title{Generalizing Shell Theorem to Constant Curvature Spaces \\ in All Dimensions and Topologies}


\author{Ava K. Tse}
\affiliation{Scripps College, Claremont, CA 91711, USA.}

\author{Olivia M. Markowich}
\affiliation{Scripps College, Claremont, CA 91711, USA.}

\author{Trung V. Phan}
\email{{tphan@natsci.claremont.edu}}
\affiliation{Department of Natural Sciences, Scripps and Pitzer Colleges, \\ Claremont Colleges Consortium, Claremont, CA 91711, USA}

\begin{abstract}
A gravitational potential has the \textit{spherical property} when the field outside any uniform spherical shell is indistinguishable from that of a point mass at the center. We present the general potentials that possess this property on constant curvature spaces, using the Euler-Poisson-Darboux identity for spherical means. Our results are consistent with known findings in flat three-dimensional space and reduce to Gurzadyan's cosmological theorem when the \textit{rescaling factor} is exactly $1$. Our approach naturally extends to \textit{nontrivial spatial topologies}.
\end{abstract}

\date{\today}

\maketitle 

\begin{figure*}[h]
    \centering
\includegraphics[width=\linewidth]{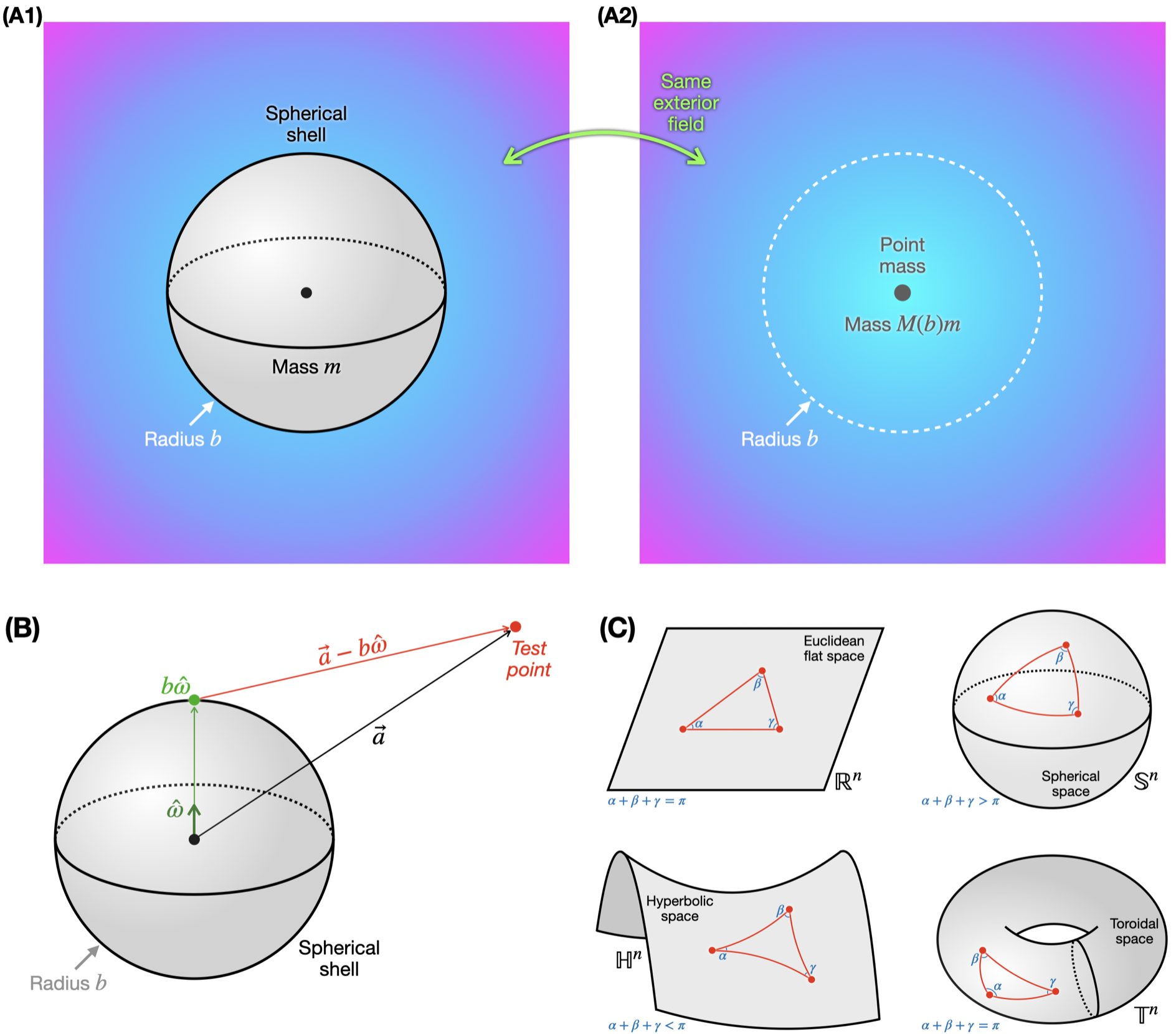}
    \caption{\textbf{Generalizing shell theorem to sphere forms in all dimensions and topologies}. Here, we focus on translationally invariant pairwise interaction potentials. \textbf{(A)} The generalized shell theorem: outside a uniform spherical shell, the field is indistinguishable from that of a central point mass, scaled by a radius-dependent factor. \textbf{(B)} The field at a \textit{test point} can be calculated by summing the contributions from all surface elements of the spherical shell. \textbf{(C)} A list of geometries we are interested in: the Euclidean flat space $\mathbb{R}^n$, the spherical space $\mathbb{S}^n$, the hyperbolic space $\mathbb{H}^n$, and the hypercubic toroidal space $\mathbb{T}^n$.}
\label{fig01}
\end{figure*}

In cosmology, Gurzadyan's theorem characterizes the full class of potentials for which a uniform spherical mass, seen from outside, is gravitationally indistinguishable from a point mass at its center \cite{gurzadyan1985cosmological,reed2022note,Carimalo2023,phan2025derivation}. This extends the \textit{{\color{black}exterior} shell theorem} in Newtonian classical gravity (point-mass equivalence for the inverse-square law) \cite{newton1833philosophiae}, first established in \cite{sneddon1949camb,kilmister1974newton} and later generalized under the name \textit{spherical property}, where the exterior field of a uniform shell coincides with that of a central point mass up to a shell-radius-dependent \textit{rescaling factor} \cite{chapman1983gravity,barnes1984gazette} (see Fig. \ref{fig01}A).\footnote{equivalently, the field outside a uniform solid sphere is indistinguishable from that of a central point mass.} These results were obtained in three-dimensional Euclidean space $\mathbb{R}^3$; here we further extend them to arbitrary $n$-dimensional constant curvature spaces, also known as spherical forms, i.e. spherical $\mathbb{S}^n$, Euclidean $\mathbb{R}^n$, and hyperbolic $\mathbb{H}^n$. This extension is made possible by linking the \textit{spherical property} to the Euler–Poisson–Darboux identity -- a connection that, to the best of our knowledge, is absent from standard references.

We start by considering the situation in flat space $\mathbb{R}^n$. Let $\phi(r)$ denote the \textit{translationally invariant and pairwise} potential as a function of the Euclidean separation $r$; then, outside a uniform spherical shell of radius $b$, the potential at a \textit{test point} located at position $\vec{a}$ away from the shell center is given by the spherical mean over all shell elements (see Fig. \ref{fig01}B):
\begin{equation}
    \Phi(\vec{a},b) = \frac{ \oint_{S^{n-1}(b)} d\sigma(\hat{\omega}) \ \phi \left(  \vec{a}-b\hat{\omega}  \right)}{\text{Vol}\left[ S^{n-1}(b)\right]} \ ,
\label{eq:full_potential}
\end{equation}
where $S^{n-1}(b)$ is the $(n-1)$-dimensional spherical surface of radius $b$, $\mathrm{Vol}[S^{n-1}(b)]$ is its total surface volume, $\hat\omega$ is a unit direction vector on the sphere, $d\sigma(\hat\omega)$ is the surface-area element, and $|\circ |$ denotes the Euclidean norm of a vector quantity $\circ$ (so $a=|\vec a|>b$ and $|\hat{\omega}|=1$). The Euler-Poisson-Darboux (EPD) identity for this spherical mean -- at least, away from singularities -- is given by \cite{john2004plane}:
\begin{equation}
    \partial_b^2 \Phi(\vec{a},b) + \frac{(n-1)}{b} \partial_b \Phi(\vec{a},b) = \Delta_{\vec{a}} \Phi(\vec{a},b) \ .
\label{eq:EPD}
\end{equation}
which is partial differential equation (PDE) that any spherical mean function $\Phi(\vec{a},b)$ should obey. The operator $\Delta_{\vec{a}} $ is the Laplacian with respect to the variable $\vec{a}$. As a demonstration, we derive this equation for $n=3$ in Appendix \ref{app:EPD}.

To satisfy the \textit{spherical property}, which specifies how the induced force behaves, we need the potential to take the form:\footnote{so that the force, given by $\vec{F}(\vec{a},b)=\vec{\nabla}_{\vec{a}} \Phi(\vec{a},b)$, satisfies $\vec{F}(\vec{a},b) = \vec{f}(\vec{a})M(b)$, where $\vec{f}(\vec{a}) = \vec{\nabla}_{\vec{a}} \phi(a) $.}
$$
    \Phi(\vec{a},b) =  \phi(a) M(b) + C(b) \ , $$
where $M(b)$ and $C(b)$ are functions to be determined;\footnote{in the limit $b \rightarrow 0$, Eq. \eqref{eq:full_potential} leads to $\Phi(\vec{a},0)=\phi(|\vec{a}|)$, and thus $M(0)=1$ and $C(0)=0$.} $M(b)$ is the \textit{rescaling factor} of the mass (as represented in Fig. \ref{fig01}A2). Place this expression in Eq. \eqref{eq:EPD} gives:
\begin{equation}
    -\lambda(b) \phi(a) - \rho(b) = \Delta_{\vec{a}} \phi(a) =  \partial_a^2 \phi(a) + \frac{(n-1)}{a} \partial_a \phi(a) 
\label{eq:solve_for_phi}
\end{equation}
where
\begin{equation}
    \lambda(b) = -\frac{\partial_b^2 M(b) + \frac{(n-1)}{b} \partial_b M(b)}{M(b)} \ \ , \ \ \rho(b) = -\frac{\partial_b^2 C(b) + \frac{(n-1)}{b} \partial_b C(b)}{M(b)} \ .
\label{eq:solve_for_M}
\end{equation}
Since the right hand side (RHS) of Eq. \eqref{eq:solve_for_phi} does not depend of $b$, the left hand side (LHS) should also not, therefore $\lambda(b)$ and $\rho(b)$ are constants, i.e. $\lambda(b)=\lambda$ and $\rho(b)=\rho$. We can then solve the PDE in Eq. \eqref{eq:solve_for_phi} to obtain the potential function $\phi(a)$ that possesses the \textit{spherical property}, which can be divided into two different cases:
\begin{itemize}
    \item \underline{$\lambda \neq 0$}: After a constant shift $$\phi(a) \rightarrow \phi(a) - \rho/\lambda \ , $$ Eq. \eqref{eq:solve_for_phi} becomes the Helmholtz equation \cite{sommerfeld1949partial}. For $\lambda > 0$, the general solution is given by a linear combination of the Bessel functions: 
    $$
        \phi(a) = A_1 a^{-(n/2-1)} J_{n/2-1}(\lambda^{1/2} a) + A_2 a^{-(n/2-1)} Y_{n/2-1}(\lambda^{1/2} a) \ ,
    $$
    where the $A$-values are arbitrary constants; and for $\lambda < 0$, it is obtained by analytic continuation (equivalently, in terms of the modified Bessel functions) \cite{abramowitz1948handbook}. When $n=3$, this expression -- after some algebraic manipulation -- can be written as
    \begin{equation}
 \phi(a) = A_3 a^{-1} \sinh(qa) + A_4 a^{-1} \cosh(qa) \ \ \text{for $\lambda = q^2 > 0$}  
 \label{eq:Yukawa}
    \end{equation}
    or
\begin{equation}
 \phi(a) = A_5 a^{-1} \sin(pa) + A_6 a^{-1} \cos(pa) \ \ \text{for $\lambda = -p^2 < 0$} \ . 
  \label{eq:osc}
    \end{equation}
The Yukawa potential, which appears in nuclear interactions (e.g. one-pion exchange) \cite{rowlinson1989yukawa}, is a special case of Eq. \eqref{eq:Yukawa}.

\item \underline{$\lambda = 0$}: This is the case when the \textit{rescale factor} is exactly $1$. Eq. \eqref{eq:solve_for_phi} becomes the Poisson equation with constant source term $\rho$, therefore the general solution for $\phi(a)$ is a linear combination between the particular solution (a quadratic function) and the fundamental solution (a Green's function, which is harmonic off the singularity):
\begin{equation}
    \phi(a) = -\frac{\rho}{2n} a^2 + A_7 a^{2-n} + A_8\ ,
\end{equation}
where the $A$-values are arbitrary constants. When $n=3$, this expression is the sum of Hookean and Coulombic potentials:
\begin{equation}
 \phi(a) = -\frac{\rho}{2n} a^2 + A_7 a^{-1} + A_8 \ . 
  \label{eq:hc}
    \end{equation}
\end{itemize}
For a sanity check, we note that  Eq. \eqref{eq:Yukawa}, Eq. \eqref{eq:osc}, and Eq. \eqref{eq:hc} agree with known results in $\mathbb{R}^3$ space \cite{chapman1983gravity,barnes1984gazette}, and Eq. \eqref{eq:hc} is the same with Gurzadyan's cosmological theorem \cite{sneddon1949camb,kilmister1974newton,gurzadyan1985cosmological,reed2022note,Carimalo2023,phan2025derivation}. To find the corresponding function $M(b)$, we solve Eq. \eqref{eq:solve_for_M} with boundary conditions $M(0)=1$ and $\partial_b M(0)=0$, which are direct consequences of Eq. \eqref{eq:full_potential}; the solutions are provided in the Appendix \ref{app:Mb_Rn}.

On curved spaces (see Fig. \ref{fig01}C), the analysis mirrors the flat case: replace the Euclidean distance with the geodesic distance and $\Delta_{\vec{a}}$ with the Laplace–Beltrami operator. The Euclidean Euler–Poisson–Darboux identity, i.e. Eq. \eqref{eq:EPD}, changes into
$$\partial_b^2 \Phi(\vec{a},b) + \frac{(n-1)}{b} \cot(b) \partial_b \Phi(\vec{a},b) = \Delta_{\vec{a}} \Phi(\vec{a},b)$$
for spherical space $\mathbb{S}^n$ (positive spatial curvature), and
$$\partial_b^2 \Phi(\vec{a},b) + \frac{(n-1)}{b} \coth(b) \partial_b \Phi(\vec{a},b) = \Delta_{\vec{a}} \Phi(\vec{a},b)$$
for hyperbolic space $\mathbb{H}^n$ (negative spatial curvature) \cite{nguyen2011spherical}. The class of potentials $\phi(a)$ with \textit{spherical property} in these spaces are solutions of an equation similar to Eq. \eqref{eq:solve_for_phi}, i.e.
\begin{equation}
    -\lambda \phi(a) -\rho = \Delta \phi(a) \ ,
\label{eq:Helmholtz_like}
\end{equation}
where $\lambda$ and $\rho$ are constants. Therefore:
\begin{itemize}
    \item In \textit{compact topology} $\mathbb{S}^n$ spaces: For \underline{$\lambda \neq 0$}, $\phi(a)$ is a linear combination of Gegenbauer functions \cite{abramowitz1948handbook}; and for \underline{$\lambda = 0$}, since local smoothness of $\phi(a)$ in this compact manifold requires $\rho=0$, $\phi(a)$ only corresponds to the fundamental solution.
    \item In \textit{open topology} $\mathbb{H}^n$ spaces: For \underline{$\lambda \neq 0$}, $\phi(a)$ is a linear combination of associated Legendre functions \cite{abramowitz1948handbook}; and for \underline{$\lambda = 0$}, $\phi(a)$ is a linear combination between the particular solution and the fundamental solution.
\end{itemize}
We show the details in Appendix \ref{app:SH}.

Observations indicate that our universe possesses spatial flatness alongside positive spacetime curvature driven by cosmic expansion \cite{aghanim2020planck}. An interesting direction for future exploration is to consider more \textit{nontrivial spatial topologies} \cite{luminet2003dodecahedral} (and also for \textit{nonspherical shell}), in which we can apply the same approach in this work. In general, the EPD identity takes the vector form -- i.e. Eq. \eqref{eq:Helmholtz_like} with the potential as a function of the full displacement $\vec{a}$ rather than only its norm:
\begin{equation}
    -\lambda \phi(\vec{a}) - \rho = \Delta \phi(\vec{a}) \ .
\label{eq:EPD_general}
\end{equation}
Because this differential relation is local, it remains unchanged; only the global geometry differs. We work out an example (for generalizing the spherical shell theorem) in flat hypercubic torus $\mathbb{T}^n$ spaces (see Fig. \ref{fig01}C) -- with common compactification length $L$ in all directions -- in Appendix \ref{app:torus}.

{\color{black}
We would like to end this brief note by emphasizing that, here, we have only scratched the surface of a deep research direction, with many open questions remaining. A natural next step is to find the most general potential $\phi(\vec a)$ that are also consistent with the \textit{interior} shell theorem in Newtonian gravity (vanishing gravitational acceleration inside a spherical shell, i.e. $\nabla_{\vec a}\Phi(\vec a,b)=0$ for $|\vec a|<b$). We conjecture that this condition can be satisfied if $\phi$ is harmonic, corresponding to Eq. \eqref{eq:EPD_general} with $\lambda=0$ and $\rho=0$ (so it already satisfies the spherical property). An example, a numerically study on $\mathbb{H}^2$ is provided in Appendix \ref{app:interior}. If this is indeed the only possible solution, then compact spaces admit no nontrivial $\phi(\vec{a})$ that yields an interior shell theorem.
}

\textbf{Acknowledgement}: We would like to thank Hieu L. D. Vu for useful discussions.

\textbf{Interest Statement}: The authors have no conflicts to disclose.

\bibliography{main}
\bibliographystyle{apsrev4-2}

\appendix 

\section{EPD identity in $\mathbb{R}^3$ \label{app:EPD}}

In $\mathbb{R}^3$, the direction unit vector can be written in geographic angles $(\theta,\varphi)$ as 
$$\hat{\omega} = (\sin\theta \cos\varphi, \sin\theta \sin\varphi, \cos\theta) \ , $$ 
and Eq. \eqref{eq:full_potential} becomes:
$$\Phi (\vec{a},b) = \frac1{4\pi} \int^{\pi}_0 d\theta \sin\theta \oint  d\varphi \ \phi(\vec{a}-b \hat{\omega}) \ . $$
Note that here we consider a general $(\vec{a}-b\hat{\omega})$-dependence, not just on the norm. Our aim is to prove the Euler-Poisson-Darboux identity:
$$ \partial_b^2 \Phi (\vec{a},b) + \frac{2}b \partial_b \Phi(\vec{a},b) = \Delta_{\vec{a}} \Phi(\vec{a},b) \ . $$
Let us define $\vec{R}=\vec{a} - b \hat{\omega}$, $\vec{r}=\vec{a}-\vec{R}$, and $r = |\vec{r}|=b$. The Laplacian $\Delta_{\vec{r}}$ in the spherical coordinates $(r,\theta,\varphi)$ centered at $\vec{a}$ is given by:
$$\Delta_{\vec{r}} = \partial_r^2 + \frac{2}r \partial_r + \frac1{r^2} \Delta_{(\theta,\varphi)} \ \ \text{where} \ \ \Delta_{(\theta,\varphi)} = \frac1{\sin\theta} \partial_\theta (\sin\theta \partial_\theta ) + \frac1{\sin^2\theta} \partial_\varphi^2 \ . $$
Due to translational invariance, we have $\Delta_{\vec{a}} \phi(\vec{a}-b\hat{\omega}) = \Delta_{\vec{R}} \phi(\vec{R})$, therefore:
\begin{equation} 
\begin{split}
    \Delta_{\vec{a}} \Phi(\vec{a},b) &= \frac1{4\pi} \int^{\pi}_0 d\theta \sin\theta \oint   d\varphi \ \left( \partial_r^2 + \frac{2}r \partial_r + \frac1{r^2} \Delta_{(\theta,\varphi)} \right) \phi(\vec{a}-b \hat{\omega})
    \\
    &= \frac1{4\pi} \int^{\pi}_0 d\theta \sin\theta \oint  d\varphi \ \left( \partial_b^2 + \frac{2}b \partial_b + \frac1{b^2} \Delta_{(\theta,\varphi)} \right) \phi(\vec{a}-b \hat{\omega}) \ .
\end{split} 
\label{eq:Lap_Phi}
\end{equation}
In the above equation, to go from the first- to the second-line we use $r=b$, e.g. any variation along the ray $\vec{r}$ is equivalent to varying $b$, hence $\partial_r = \partial_b$ and $\partial_r^2 = \partial_b^2$.

For the angular Laplacian $\Delta_{(\theta,\varphi)}$ acting on $\phi(\vec{a}-b \hat{\omega})$, we can treat it as a function of the geographic angle variables $(\theta,\varphi)$, i.e. $\phi(\vec{a}-b \hat{\omega}) = f(\theta,\varphi)$. Since Eq. \eqref{eq:Lap_Phi} can be separated into:
\begin{equation}
\Delta_{\vec{a}}\Phi(\vec{a},b) = \left( \partial_b^2 + \frac{2}b \partial_b \right)  \Phi(\vec{a},b) + I \ ,
\label{eq:proto_EPD}
\end{equation}
the integration part $I$ can be written as:
\begin{equation}
\begin{split}
    I &= \frac1{4\pi b^2} \int^{\pi}_0 d\theta \sin\theta \oint  d\varphi \ \Delta_{(\theta,\varphi)} f(\theta,\varphi) 
    \\
    &= \frac1{4\pi b^2} \left\{ \oint d\phi \int^{\pi}_0 d\theta \ \partial_\theta \left[ \sin\theta \partial_\theta f(\theta,\varphi) \right] + \int^\pi_0 \frac{d\theta}{\sin\theta} \oint d\phi \  \partial_\phi^2 f(\theta,\varphi) \right\}
\end{split}
\end{equation}
The second term here is trivially $0$, while the first term after the $\int d\theta$ integration gives 
$$\left[ \sin\theta \partial_\theta f(\theta,\varphi) \right]\Big|^{\pi}_{0} \ , $$
which is equal to $0$ since $\sin\theta = 0$ for both $\theta=0$ and $\theta=\pi$. Together, $I=0$, and what are left of Eq. \eqref{eq:proto_EPD} is exactly the EPD identity.

\section{The \textit{rescaling factor} $M(b)$ in $\mathbb{R}^n$ \label{app:Mb_Rn}}

We want to find the function $M(b)$ that satisfies the boundary conditions $M(0)=1$ and $\partial_b M(0)=0$ while solving the following differential equation:
\begin{equation}
    \partial_b^2 M(b) + \frac{n-1}{b} \partial_b M(b) + \lambda M(b) = 0 \ ,
\end{equation}
which is identical to Eq. \eqref{eq:solve_for_M} up to simple algebraic rearrangement. Here we present the solutions for different values of $\lambda$:
\begin{itemize}
    \item \underline{$\lambda = 0$}: The \textit{rescaling factor} is the identity, i.e. $M(b)=1$.
    \item \underline{$\lambda > 0$}: Define $\lambda = q^2 > 0$, then the \textit{rescaling factor} is given by:
    \begin{equation}
        M(b)= \Gamma(n/2) \left(2/qb \right)^{n/2 - 1} J_{n/2 - 1} (qb) \ .
    \end{equation}
    When $n=3$, this expression becomes $M(b)=\sin(qb)/qb$.
    \item \underline{$\lambda < 0$}: Define $\lambda = -p^2 < 0$, then the \textit{rescaling factor} can be written using modified Bessel function:
    \begin{equation}
        M(b)= \Gamma(n/2) \left(2/qb \right)^{n/2 - 1} I_{n/2 - 1} (qb) \ .
    \end{equation}
    When $n=3$, this expression becomes $M(b)=\sinh(pb)/pb$.
\end{itemize}

\section{Potentials possess \textit{spherical property} in $\mathbb{S}^n$ and $\mathbb{H}^n$ \label{app:SH}}

In this Appendix, the $A$-values are arbitrary constants.

Consider $\mathbb{S}^n$ spaces with curvature $\kappa=k^2 >0$. For \underline{$\lambda \neq 0$}, $\phi(a)$ is a linear combination of Gegenbauer functions \cite{abramowitz1948handbook}:
    $$ \phi(a) = A_9 C^\alpha_\nu \left[ \cos(ka)\right] + A_{10} Q^\alpha_\nu \left[ \cos(ka)\right] - \rho/\lambda  \ , $$
    in which 
    $$\alpha = (n-1)/2 \ \ , \ \ \nu (\nu + n-1) = \lambda /k^2 \ .$$
    For \underline{$\lambda = 0$}, local smoothness of $\phi(a)$ in this compact manifold requires $\rho=0$,\footnote{since we always have $\int_{\mathbb{S}^n} d^n\vec{a} \  \Delta_{\vec{a}} \phi(a) = 0$ (properties on a compact manifold), thus from $\Delta_{\vec{a}} \phi(a)=-\rho$ we obtain  $\int_{\mathbb{S}^n} d^n\vec{a} \ (-\rho) = -\rho \text{Vol}(\mathbb{S}^n)=0$, which leads to $\rho=0$.} therefore $\phi(a)$ corresponds to a fundamental solution:
    $$ \phi(a) = A_{11} \left[ 2\sin(ka/2) \right]^{2-n} + A_{12} \ .$$

Next, we look at $\mathbb{H}^n$ spaces with curvature $\kappa=-l^2 <0$: For \underline{$\lambda \neq 0$}, $\phi(a)$ is a linear combination of associated Legendre functions \cite{abramowitz1948handbook}:
    $$ \phi(a) = A_{13} P^{-\alpha}_{-1/2+i\tau} \left[ \cosh(la)\right] + A_{14} Q^{-\alpha}_{-1/2+i\tau} \left[ \cosh(la)\right] - \rho/\lambda  \ , $$
    in which 
    $$\alpha = (n-1)/2 \ \ , \ \ \tau^2 = \lambda /l^2 - \alpha^2 \ .$$ 
    For \underline{$\lambda = 0$}, $\phi(a)$ can be found as a linear combination between the particular solution (regular) and the fundamental solution (singular):
    $$ \phi(a) = \rho \int_0^a da' \ \frac{\int^{a'}_0 da'' \ S(a'')}{S(a')} + A_{15} \int^\infty_a \frac{da'}{S(a')} + A_{16} \ , $$
    where we define $S(a)=\left[ \sinh(la)\right]^{n-1}$. When $n=3$, this admits a closed form expression:
    $$ \phi(a) = \frac{\rho}{2 l} a \coth(la) + A_{15} \left[\coth (la)  -1 \right] + A_{16} \ . $$

\section{Potentials possess \textit{spherical property} in $\mathbb{T}^n$  \label{app:torus}}

Consider hypercubic $\mathbb{T}^n$ spaces with a common compactification length $L$. We want to find to general solution $\phi(\vec{a})$ of Eq. \eqref{eq:EPD_general}.

For \underline{$\lambda \neq 0$}, $\phi(\vec{a})$ can only be the constant function $-\rho/\lambda$ (so no force) unless $\lambda$ is an eigenvalue of $\Delta_{\vec{a}}$, i.e. $\lambda = (2\pi |\vec{m}|/L)^2$ with $\vec{m} \in \mathbb{Z}^n \setminus \{ 0\}$. When this is the case, the most general potential is of the form:  
\begin{equation}
    \phi(\vec{a}) =  c_{\vec{m}} \cos\left[2\pi (\vec{m}.\vec{a})/L\right] + s_{\vec{m}} \sin\left[2\pi (\vec{m}.\vec{a})/L\right] - \rho/\lambda \ ,
\end{equation}
where $c_{\vec{m}}$ and $s_{\vec{m}}$ are arbitrary constants.

For \underline{$\lambda = 0$}, similar to the argument made in Appendix \ref{app:SH} for $\mathbb{S}^n$ spaces, local smoothness of $\phi(\vec{a})$ in this compact manifold (or any compact manifold) requires $\rho=0$, therefore $\phi(\vec{a})$ corresponds to a fundamental solution:
    $$ \phi(a) = B_{1} \sum_{\vec{m} \in \mathbb{Z}^n \setminus \{ 0\}} \frac{\exp\left[ i2\pi(\vec{m}.\vec{a})/L\right]}{ (2\pi|\vec{m}|/L)^2} + B_{2} \ , $$
where $i$ is the unit imaginary number and the $B$-values are arbitrary constants.

{\color{black}
\section{On Potentials satisfy the interior shell theorem \label{app:interior}}

To ensure vanishing gravitational acceleration inside the spherical shell, we require that the spherical mean $\Phi(\vec{a},b)$, calculated from $\phi(\vec{a})$ via Eq. \eqref{eq:full_potential}, be independent of $\vec{a}$ for any interior point i.e. $|\vec{a}|<b$. Thus, $\Delta_{\vec{a}} \Phi(\vec{a},b)=0$ and therefore:
\begin{equation} \oint_{S^{n-1}(b)} d\sigma(\hat{\omega}) \ \Delta \phi \left(  \vec{a}-b\hat{\omega}  \right) = 0 \ . 
\label{eq:to_0}
\end{equation}
If the potential is harmonic i.e. $\Delta \phi = 0$, then this integral is trivially satisfied. This is a special case of Eq. \eqref{eq:EPD_general}, in which $\lambda=0$ and $\rho=0$, so a harmonic potential not only satisfies the spherical property but also might results in the interior shell theorem. We therefore conjecture that harmonicity is the necessary condition for all of these properties to hold.


\begin{figure*}[h]
    \centering
\includegraphics[width=\linewidth]{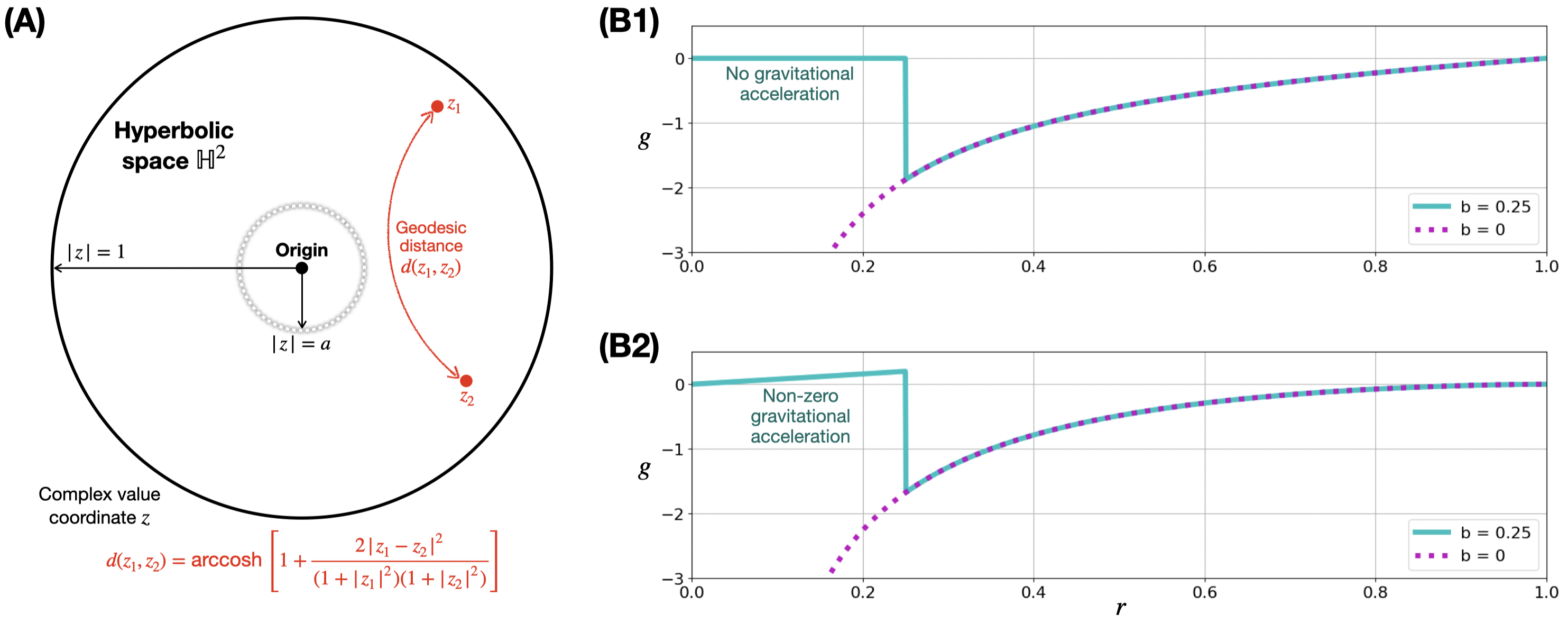}
    \caption{\textbf{Shell theorems in hyperbolic $\mathbb{H}^2$ space}. We consider the generalized exterior shell theorem -- the spherical property -- and the interior shell theorem by investigating the gravitational acceleration via Eq. \eqref{eq:gra_acc}. \textbf{(A)} This space in a complex coordinate system. \textbf{(B1)} A potential possesses spherical property but fails interior shell theorem. \textbf{(B2)} A potential possesses both spherical property and interior shell theorem.}
\label{fig02}
\end{figure*} 

We already know this is true for flat spaces, so let us explore this conjecture numerically in other spaces. Consider the $\mathbb{H}^2$ space, in the unit-disk complex coordinate (see Fig. \ref{fig02}A), with the followings:
\begin{itemize}
    \item A potential satisfying Eq. \eqref{eq:EPD_general} with $\lambda = 0$ and $\rho = 0$  can obey both the spherical property and the interior shell theorem (see Fig. \ref{fig02}B1).
    \item A potential satisfying Eq. \eqref{eq:EPD_general} with $\lambda \neq 0$ and $\rho \neq 0$ can obey the spherical property, yet fails the interior shell theorem (see Fig. \ref{fig02}B2).
\end{itemize}
Here, we work with a hyperbolic space with curvature $\kappa=-1$ associated with the metric:
$$ds^2 = \frac{4|dz|^2}{(1-|z|^2)^2} \ .$$
The geodesics in this coordinate system are circular arcs, whose distance between two locations $z_1$ and $z_2$ can be calculated via:
\begin{equation}
    d(z_1,d_2) = \text{arccosh}\left[ 1 + \frac{2|z_1-z_2|^2}{(1+|z_1|^2)(1+|z_2|^2)} \right] \ .
\end{equation}
For Fig. \ref{fig01}B1, we consider a harmonic potential
$$ \phi(\vec{a}) = -\ln\left[ \tanh (a/2)\right] \ . $$
For Fig. \ref{fig01}B2, we consider the potential 
$$\phi(\vec{a}) = 1 - Q_\nu \left[\cosh(a) \right] \ \ \text{with} \ \ \nu = \frac{\sqrt{5}-1}{2} \ ,  $$
which corresponds to $\lambda=1$ and $\rho=-1$.
The gravitational acceleration at a given position $z$ is calculated from:
\begin{equation}
    g = \frac{\vec{a}}{a}.\left[ -\vec{\nabla}_{\vec{a}} \Phi(\vec{a},b) \right] = -\left(\frac{1-r^2}2 \right) \partial_r \Phi(r,b) \ ,
    \label{eq:gra_acc}
\end{equation}
where we represent a position $\vec{a}$ with a complex value $z$ and $r = |z|$. Note that we have one rescaling parameter, corresponding to $M(b) \neq 1$.

If we ask only for a potential $\phi(\vec{a})$ consistent with the interior shell theorem, without additionally imposing the spherical property, then it is not difficult to guess a solution, at least in flat $\mathbb{R}^n$ spaces. For instance, requiring $\Phi(\vec{a},b)=0$ throughout the interior of a sphere leads to the explicit examples:
$$ \phi(\vec{a}) = \frac{a_1}{|\vec{a}|^2} \ \ \text{for $n=2$} \ , $$
and
$$ \phi(\vec{a}) = \frac{(2-n)a_1}{|\vec{a}|^n} \ \ \text{for $n\geq 3$} \ , $$
where $a_1$ denotes a chosen Cartesian component of $\vec{a}$. These potentials do not have rotational symmetry.

}


\end{document}